\documentstyle[12pt,graphicx]{article}
\begin{document}
\title{Gravitating Magnetic Monopole in the Global Monopole Spacetime}
\author{J. Spinelly\thanks{E-mail: spinelly@fisica.ufpb.br},\\
U. de Freitas\thanks{E-mail: umbelino@ccen.ufpb.br}\\ and 
E. R. Bezerra de Mello \thanks{E-mail: emello@fisica.ufpb.br}\\
Departamento de F\'{\i}sica-CCEN\\
Universidade Federal da Para\'{\i}ba\\
58.059-970, J. Pessoa, PB\\
C. Postal 5.008\\
Brazil}
\maketitle
\begin{abstract}
In this paper we study the regular self-gravitating 't Hooft-Polyakov magnetic
monopole in a global monopole spacetime. We show that for the large distance,
the structure of the manifold corresponds to the Reissner-Nordstr\"{o}m
spacetime with a solid angle deficit factor. Although we analyze static and
spherically symmetric solutions, it is not possible to solve analytically
the system of coupled differential equations and only numerical evaluations 
can provide detailed information about the behavior of this system at the 
neighborhood of the defect's core. So, for this reason we solve numerically 
the set of differential equations for the metric tensor and for the matter 
fields for different values of the Higgs field vacuum expectation value, 
$\eta$, and the self-coupling constant, $\lambda$.
\end{abstract}

\newpage
\renewcommand{\thesection}{\arabic{section}.}
\section{Introduction}
$        $

The self-gravitating 't Hooft-Polyakov magnetic monopole \cite{HP} in a curved
spacetime has been studied a few years ago considering it as a magnetic
point charge \cite{Bais, Cho}. The exact solution obtained for the metric 
tensor has the Reissner-Nordstr\"{o}m form corresponding to a point (magnetic)
charge $g=1/e$. A regular solution for this system has been presented by
Nieuwenhuizen {\it et al} \cite{Nieuw}. There, they have constructed a
positive-definite functional energy function of the matter fields only.
They claimed that this is enough to prove the existence of non-singular 
monopole solutions. More recently Lee {\it et al} \cite{Lee} and Ortiz
\cite{Ortiz}  have shown that non-singular monopole solutions exist only if 
the Higgs vacuum expectation value, $\eta$, is smaller or equal to a critical 
value, $\eta_{cr}$, which is of order of the Planck mass. In the limiting 
case the monopole becomes a black hole, with the region outside the horizon 
described by the Reissner-Nordstr\"{o}m solution.

Barriola and Vilenkin \cite{BV} have analyzed the effect in the geometry
of the spacetime produced by a system composed by Higgs field only, which
undergoes to a spontaneous breakdown of global $O(3)$ gauge symmetry.
They noticed that the solution for the metric tensor is similar to the
Schwarzchild spacetime with an additional solid angle deficit and a nonzero 
scalar curvature. They pointed out that for large value of the
geometric mass, the model describes a black hole carrying a global monopole.

One of the main differences between the large distance behaviors in the 
geometries of the spacetime produced by both topological defects, the local 
and the global monopoles, is due to their energy densities, which for the 
global monopole case decreases as $1/r^2$. This behavior 
is responsible for the solid angle deficit presented by this geometry.

In this paper we continue the discussion related with this
topic and consider both type of topological defects in the same
model. We analyze the effects produced by local and global monopole on the
geometry of the manifold. We investigate the possibility
of this system to present regular solution and we also analyze its behavior
near and far away from the defect's core. So the basic idea of this model
is to describe a regular topological defect which presents a magnetic field
besides to present a solid angle deficit.

Assuming the existence of such object in a typical galaxy, the total energy
contained inside it would be strongly provided by the global Higgs
field \cite{Salgado} \footnote{In fact, the energy density outside the 
global monopole is $T^t_t \approx -\eta^2/r^2$, consequently the total 
Newtonian mass inside a space region of radial extent $R$ is $E(R)\approx 
4\pi G\eta^2R$. Considering $R$ the typical radius of a galaxy $R\approx 
15Kpc$ and for the symmetry breaking scale $\eta \approx 10^{16} GeV$ which 
is the value for grand unified theories, this total energy due to the global 
monopole is approximately ten times the mass of the galaxy.}. Astrophysics
bounds on the flux of magnetic monopole and evidence that the galactic 
magnetic field is mainly azimuthal \cite{Parker} indicate that the excess
number density of such object, if they really exist, is very small. Moreover, 
upper bounds on the number density of global monopole is at most one in the
local group of galaxies as pointed out by Hiscock \cite{Hiscock}. 

Differently from a pure global monopole, this compost topological defect
exerts a gravitational interaction on surrounding matter, apart from the
electromagnetic one on charged particles. So, such object shares with both,
global and magnetic monopole, some of their relevant properties. Numerical
simulation related with the upper bound on the number density of them in the
Universe, may be developed in a similar manner as it was developed to
global monopole only in the paper by Bennett and Rhie \cite{BE}.

The complete information about this system requires the knowledge of the
behaviors of the matter and gravitational fields, i.e., we have to know how 
these fields change along the distance and how they are connected; besides we 
also want to know how these fields' behaviors are affected when the energy 
scale of breakdown of gauge symmetry and the Higgs self-coupling are varying. 
Because it is impossible to solve analytically the complete set of coupled 
differential equation associated with this system, only numerical 
analysis makes possible to obtain these informations. Numerical analysis
of self-gravitating magnetic monopole has been developed by several
authors, see for example Refs. \cite{Lee} and \cite{Peter}. For the global 
monopole case Harari and Loust\'{o} \cite{Lousto} have shown numerically the 
behavior of the Higgs field and how it is affected by the variation of  the 
parameter $\eta$.  More recently Maison \cite{Maison} and Liebling 
\cite{Liebling} have analyzed the stability condition for the global monopole 
solution. They found that for $\eta$ bigger than some critical value, the 
global monopole fail to be static.

This paper is organized as follows. In section $2$ we briefly review some
of the relevant characteristics of the local and global monopole in a 
curved spacetime. We also introduce the model used to describe the system
which presents the topological defect formed by both monopoles and derive
the equations of motion which governs the behavior of this object.
Because it is impossible to solve analytically this set of differential 
equations we leave for the section $3$ its numerical analysis. From our
results it is possible to exhibit the behaviors for matter and gravitational
fields, their dependence with the distance from the point to the monopole's 
core and how they are connected among other pertinent informations. In
section $4$ we present our conclusions and some important remarks about
this paper.

\section{Field Equation for the Compost Topological Defect}

In this section we introduce the model which, by a spontaneous breakdown of 
gauge symmetry, gives rise to a non-Abelian magnetic and global monopoles in a 
curved spacetime. This defect presents both properties of its
constituent: a magnetic field and a solid angle deficit.  Below we shall
briefly review both topological defects separately.

The global monopole is a defect obtained by a system composed by a 
self-coupling Higgs isotriplet field which undergoes to a spontaneous breakdown
of global $O(3)$ gauge symmetry to $U(1)$. Coupling this matter field with 
the Einstein equation, a spherically symmetric metric tensor given by the
line element 
\begin{equation}
\label{S}
ds^2=-B(r)dt^2+A(r)dr^2+r^2(d\theta^2+\sin\theta^2d\phi^2) \ ,
\end{equation}
presents regular solutions for the radial functions $B(r)$ and $A(r)$, that
for points far from the monopole's core are given by \cite{BV}
\begin{equation}
B(r)=A(r)^{-1}=1-8\pi G\eta^2-2GM/r \ ,
\end{equation}
$\eta$ being the scale energy where the symmetry is broken. The parameter $M$ 
is approximately the mass of the monopole. Neglecting the mass term and 
rescaling  the time variable, we can rewrite the monopole metric tensor as
\begin{equation}
\label{GM}
ds^2=-dt^2+\frac{dr^2}{\alpha^2}+r^2(d\theta^2+\sin\theta^2d\phi^2) \ ,
\end{equation}
where the parameter $\alpha^2=1-8\pi G\eta^2$ is smaller than unity. The
above geometry presents no Newtonian potential, it is not flat\footnote{The 
scalar curvature associated with this spacetime is $R=\frac{2(1-
\alpha^2)}{r^2}$.} and the solid angle of a sphere of unity radius is
$4\pi\alpha^2$, so smaller than $4\pi$.

The energy-momentum tensor associated with the matter field, outside the 
monopole's core can be approximately written by
\begin{equation}
T^t_t\approx T^r_r \approx -\frac{\eta^2}{r^2} \ ,\ T^\theta_\theta
\approx T^\phi_\phi \approx 0 \ ,
\end{equation}
consequently the energy is linearly divergent at large distance: $E(r)\approx 
4\pi G\eta^2 r$. 

The magnetic monopole is also a topological defect described by a system 
composed by a self-coupling Higgs isotriplet field which interacts with a 
Yang-Mills gauge field. This system presents a local $SO(3)$ gauge symmetry 
which is spontaneously broken down to $U(1)$. In a flat spacetime this theory 
gives rise to 't Hooft-Polyakov monopole with  magnetic charge and finite 
energy \cite{HP}. This system has been firstly analyzed in a curved spacetime 
in Refs. \cite{Bais, Cho}. In these papers the authors have shown that
this system presents as an exact solution a metric tensor identical with
the Reissner-Nordstr\"{o}m one
\begin{equation}
B(r)=\frac1{A(r)}=1-\frac{2GM}r+\frac{4\pi G}{e^2r^2} \ ,
\end{equation}
where $M$ ia a constant of integration, identified as the mass of the
monopole and $1/e$ is its magnetic charge. 

The energy-momentum tensor associated with the matter fields compatible
with this singular solution is
\begin{equation}
T^t_t=T^r_r=-T^\theta_\theta=-T^\phi_\phi=-\frac1{2e^2r^4} \ .
\end{equation}

In their remarkable paper Nieuwenhuizen {\it et al} \cite{Nieuw} have 
proved the existence of non-singular self-gravitating magnetic monopole.
In order to do that, they have constructed a positive-definite functional 
energy whose minimum value was claimed to be attained by a stable non-singular 
solution. They also have presented the boundary conditions obeyed by regular 
solutions at the monopole's core and show that the asymptotic form of the 
metric tensor is a Reissner-Nordstr\"{o}m geometry. More recently Lee 
{\it et al} \cite{Lee} and Ortiz \cite{Ortiz} have analyzed again the 
self-gravitating magnetic monopole system and observed that for very heavy 
monopole there is no non-singular solutions. They pointed out that when 
$G\eta^2$ becomes large $A^{-1}(r)$ presents a local minimum which approaches 
to zero; so for some critical value there appears a horizon and the monopole 
becomes a black-hole with the region outside to the horizon described by the 
Reissner-Nordstr\"{o}m metric spacetime. They present numerical solutions
for the matter and gravitational fields for different values of the parameter
$\eta^2$, where explicitly the horizon shows up.

After this brief review let us introduce the model proposed by us. The basic 
idea of this model is to describe both topological defects on the same time.
In order to do that we endow this model with a gauge group product of
two different gauge groups symmetry. Because we want to obtain magnetic 
monopole configuration we have to gauge one of them. Also we introduce two
 Higgs fields in $({\bf 3,1})$ and $({\bf 1,3})$ representations of the 
$G:=SO_L(3)\otimes O_G(3)$ groups, where the subindices refer to local 
$(L)$ and global $(G)$ gauge symmetries. The Higgs fields are responsible 
for the spontaneous break of gauge symmetries $SO_L(3)\otimes O_G(3)$ 
to $U_L(1)\otimes U_G(1)$. Moreover in order to simplify our analysis
we shall consider two situations: The first case is obtained taking the Higgs 
self-coupling constants and vacuum expectation values the same in both 
sectors and do not admit a direct coupling between them. The second case
is a particular situation of the first one taking the self-coupling associated 
with the local sector vanishing. In the latter, the system also presents
localized self-gravitating magnetic monopole. The Lagrangian density which 
governs the more general case, i.e., the first one, is:
\begin{equation}
\label{L}
{\cal L}_M=-\frac14(F^a_{\mu\nu})^2-\frac12g^{\mu\nu}(D_\mu\phi^a)
(D_\nu\phi^a)-\frac12g^{\mu\nu}(\partial_\mu \chi^a)(\partial_\nu\chi^a)-
V(\phi^a,\chi^a) \ ,
\end{equation}
with the Latin indices referring to the internal gauge groups $a,\  b=1, \ 2, 
\ 3$. We also have
\begin{equation}
D_\mu\phi^a=\partial_\mu\phi^a-e\epsilon_{abc}A_\mu^b\phi^c \ ,
\end{equation}
\begin{equation}
F_{\mu\nu}^a=\partial_\mu A_\nu^a-\partial_\nu A_\nu^a-e\epsilon_{abc}
A_\mu^b A_\nu^c \ ,
\end{equation}
and
\begin{equation}
V(\phi^a,\chi^a)=\frac{\lambda}4\left(\phi^a\phi^a-\eta^2\right)^2+
\frac{\lambda}4\left(\chi^a\chi^a-\eta^2\right)^2 \ .
\end{equation}

In the following analysis we shall consider only static spherically 
symmetric solutions, for this reason the metric tensor is written in the
form presented by (\ref{S}).

The ansatz adopted to describe both topological defects is the usual one in
flat spacetime written in terms of 'Cartesian' coordinates as
\begin{equation}
\label{Higgs1}
\chi^a(x)=\eta f(r)\hat{x}^a \ ,
\end{equation}
\begin{equation}
\phi^a(x)=\eta h(r)\hat{x}^a \ ,
\end{equation}
\begin{equation}
A_i^a(x)=\epsilon_{iaj}\hat{x}^j\frac{1-u(r)}{er} \ ,
\end{equation}
and
\begin{equation}
\label{A0}
A_0^a(x)=0 \ .
\end{equation}

Because we are seeking static solutions all properties of the system may be
described by the Lagrangian which is the sum of the Einstein one, $L_E$, and
the covariant matter Lagrangian, $L_M$:
\begin{equation}
L_E=\frac1{16\pi G}\int d^3x \ \sqrt{-g}\ R 
\end{equation}
and
\begin{equation}
L_M=\int d^3x\ \sqrt{-g}\ {\cal L}_M \ .
\end{equation}

Substituting the configurations (\ref{Higgs1}) - (\ref{A0}) into (\ref{L}),
together with the spherically symmetric metric tensor (\ref{S}), we obtain
for the matter field the following Lagrangian:
\begin{equation}
\label{LM}
L_M=-4\pi\int_0^\infty dr \ r^2 \ \sqrt{AB}\left[\frac{{\cal K}(f,h,u)}A+
{\cal U}(f,h,u)\right] \ ,
\end{equation}
where
\begin{equation}
\label{K}
{\cal K}(f,h,u)=\frac12\eta^2(f')^2+\frac12\eta^2(h')^2+\frac{(u')^2}
{e^2r^2} \ ,
\end{equation}
and
\begin{equation}
\label{U}
{\cal U}(f,h,u)=\frac{(u^2-1)^2}{2e^2r^4}+\frac{\eta^2 u^2 h^2}{r^2}+
\frac{\eta^2f^2}{r^2}+\frac{\lambda\eta^4}4(h^2-1)^2+\frac{\lambda\eta^4}4
(f^2-1)^2 \ ,
\end{equation}
where the primes  denote differentiation with respect to $r$. 

The Einstein Lagrangian for the metric tensor (\ref{S}) reads
\begin{eqnarray}
L_E&=&\frac1{4G}\int_0^\infty dr\ \left[-\frac1{\sqrt{AB}}\left(r^2B'\right)'+
\frac{r^2B'A'}{2A\sqrt{AB}}+\frac{r^2(B')^2}{2B\sqrt{AB}}+\right.
\nonumber\\
&&\left.\frac{2rA'}A\sqrt{\frac B A}+2\sqrt{AB}\left(1-
\frac1A\right)\right] \ .
\end{eqnarray}

Following the procedure adopted in \cite{Nieuw} it is possible to work
with the Lagrangian below, $L_E'$, which differs from the previous one 
by a total derivative:
\begin{equation}
\label{LE}
L_E'=\frac1{4 G}\int_0^\infty \ dr\ r \sqrt{AB}\left(\frac1A-1\right)
\left(\frac{A'}A+\frac{B'}B\right) \ .
\end{equation}

The total Lagrangian for this system can be given as the sum of (\ref{LM})
with (\ref{LE}). As we can see (\ref{LE}) can be written in terms of two
new radial fields $X=\sqrt{AB}$ and $Y=\sqrt{B/A}$. The Euler-Lagrange equations 
for the gravitational degrees of freedom can be obtained by:
\begin{equation}
\label{Y}
\left(rY\right)^{'}=X\left(1-8\pi G r^2{\cal U}\right)
\end{equation}
and
\begin{equation}
\label{X}
\frac{X'}X=8\pi Gr{\cal K} \ .
\end{equation}

Integrating (\ref{X}) assuming that at infinity $X(\infty)=1$ we obtain
\begin{equation}
\label{A}
A(r)=\frac1{B(r)}\exp\left[16\pi G\int_\infty^rdr^{'}\ r^{'}\ {\cal K}(f,h,u)
\right] \ .
\end{equation}

Now going back to (\ref{Y}) we obtain
\begin{equation}
\label{A2}
\left(\frac r A\right)'=1-8\pi Gr^2\left[\frac{{\cal K}(f,h,u)}A+
{\cal U}(f,h,u)\right] \ .
\end{equation}

Integrating the above equation assuming the regularity condition on $r/A(r)$ at
origin, we have
\begin{equation}
\label{A1}
\frac1{A(r)}=\alpha^2-\frac{2GM(r)}r \ ,
\end{equation}
being $\alpha^2=1-8\pi G\eta^2$ and $M(r)$ given by the integral
\begin{eqnarray}
\label{M}
M(r)&=&4\pi\int_0^r dr'\ r'^2\left[\frac{\eta^2}{2A}\left[(f')^2+(h')^2\right]
+\frac{(u')^2}{Ae^2{r^{'}}^2}+\frac{\eta^2u^2h^2}{{r^{'}}^2}+\eta^2\frac{(f^2-1)}{{r^{'}}^2}
\right.
\nonumber\\
&&\left.+\frac{(u^2-1)^2}{2e^2{r^{'}}^4}+\frac{\lambda\eta^4}4(h^2-1)^2+\frac{\lambda\eta^4}4(f^2-1)^2
\right] \ .
\end{eqnarray}

We can also rewrite $M(r)$ in a different way substituting (\ref{A1}) into the
right hand side of (\ref{A2}). After some steps we get
\begin{equation}
M(r)=\exp\left[-\int_0^r dr'\ p(r')\right]\int_0^r dr'\ q(r')\ \exp\left[
\int_0^{r'} dr''\ p(r'')\right] \ ,
\end{equation}
with
\begin{equation}
p(r)=8\pi Gr{\cal K}(f,h,u)
\end{equation}
and
\begin{equation}
q(r)=4\pi \left[r^2\left(\alpha^2{\cal K}(f,h,u)+{\cal U}(f,h,u)\right)-
\eta^2\right] \ .
\end{equation}
With this procedure we have removed the factor $1/A(r)$ in the integral 
definition of $M(r)$ given in (\ref{M}). From the above equations
we can obtain the total mass written as
\begin{equation}
{\cal M}=M(\infty)=4\pi\int_0^\infty\left[r^2\left({\cal U}(f,h,u)+
\alpha^2{\cal K}(f,h,u)\right)-\eta^2\right]e^{-P(r)} \ ,
\end{equation}
where
\begin{equation}
P(r)=8\pi G\int_r^\infty dr'\ r'\ {\cal K}(f,h,u) \ ,
\end{equation}
which is a positive-definite quantity.

The gravitational field equations, (\ref{Y}) and (\ref{X}) can be rewritten 
in a different way in terms of the radial function $M(r)$ as follows:
\begin{equation}
\label{AB}
\frac{(AB)'}{AB}=16\pi Gr{\cal K}(f,h,u)
\end{equation}
and
\begin{equation}
M'(r)+8\pi Gr{\cal K}M(r)=4\pi r^2\left({\cal U}+\alpha^2{\cal K}\right)-
4\pi\eta^2 \ .
\end{equation}

For the matter fields we have:
\begin{equation}
\label{U1}
\frac1{\sqrt{AB}}\left[\sqrt{\frac BA}u'\right]'=\frac{e^2r^2}2\frac{\partial
{\cal U}}{\partial u}=\frac{u(u^2-1)}{r^2}+\eta^2e^2h^2u \ ,
\end{equation}
\begin{equation}
\label{H1}
\frac1{r^2\sqrt{AB}}\left[r^2\sqrt{\frac BA}h'\right]'=\frac1{\eta^2}\frac
{\partial {\cal U}}{\partial h}=\frac{2hu^2}{r^2}+\lambda\eta^2 h(h^2-1) \ ,
\end{equation}
and
\begin{equation}
\label{F}
\frac1{r^2\sqrt{AB}}\left[r^2\sqrt{\frac BA}f'\right]'=\frac1{\eta^2}
\frac{\partial {\cal U}}{\partial f}=\frac{2f}{r^2}+\lambda\eta^2f(f^2-1) \ .
\end{equation}

From these set of differential equations it is possible to observe that there 
is no direct interaction between the global Higgs field expressed in terms of
$f(r)$ with the magnetic sector represented by $h(r)$ and $u(r)$. However the
gravitational field interacts with both sectors. Moreover these equations
are invariant under the discrete symmetries $f \rightarrow-f$, $h \rightarrow
-h$ and $u \rightarrow -u$. The first two transformations correspond to
specific choice of monopoles configurations and the last one corresponds to a 
gauge transformation.

In order to analyze this set of differential equations let us first discuss 
the boundary conditions obeyed by the fields.

The boundary condition on the matter fields at infinity follows by the 
requirement of the topological defect be localized
\begin{equation}
f \rightarrow \pm 1, \ h \rightarrow \pm 1\  and \ u \rightarrow 0 \ .
\end{equation} 

Due to the presence of the global Higgs sector, the metric components do not
asymptote to unity. So according to the results exhibited
for the purely global monopole spacetime we can write the following boundary
conditions at infinity:
\begin{equation}
AB \rightarrow 1\ and \ M/r \rightarrow 0 \ .
\end{equation}

The last two conditions above follows from the previous one obeyed by the 
matter fields, as can be easily observed by the expressions (\ref{K}) and 
(\ref{U}). The double sign which appear for the behavior of $h$ and 
$f$ at infinity corresponds to the monopole or anti-monopole configurations. 
In this paper we shall adopt the positive sign for both Higgs fields.

The boundary conditions at origin required by regularity of our
solutions are:
\begin{equation}
\label{O}
u\rightarrow 1,\ f\rightarrow 0,\ h\rightarrow 0,\ AB\rightarrow 1,\ 
and\ M\rightarrow -4\pi\eta^2 r \ .
\end{equation}

Being satisfied these conditions, the behavior of the integrand for the 
Lagrangian associated with the matter and gravitational fields, 
Eqs. (\ref{LM}) and (\ref{LE}), vanish at origin. 

As it was pointed out in Refs. \cite{Bais, Cho}, the differential equations
obeyed by the matter fields associated with the local monopole sector only, 
admit exact (singular) solution $u=0$ and $h=1$ everywhere. However as to the 
global monopole sector, the field $f$ goes to unity only at infinity.

So, unfortunately the complete set of differential equations does not admit 
closed solution, even singular one. So the relevant aspects about this
compost defect can only be observed numerically. We leave this analysis for 
the next section. Before to end the present section we would like to make
two comments about this model:\\
$a)$ The first one refers to the positive-definite functional
energy property enjoyed by this model. In fact, eliminating the gravitational
degrees of freedom from the total Lagrangian, $L_T=L_M+L_E'$, by using 
(\ref{X}), we obtain a energy-functional, $E=-L_T$, expressed in terms of the 
matter fields as:
\begin{equation}
E=\int d^3x\ \sqrt{-g}\left({\cal U}+{\cal K}\right) \ .
\end{equation} 
\noindent
$b)$ The second point that we want to mention is that a pointlike
topological defect which takes into account a (point) magnetic charge
$g=1/e$ in a solid angle deficit geometry, can be obtained by considering
a non-dynamical energy-momentum tensor below in the Einstein equation:
\begin{equation}
T^t_t=T^r_r=-\frac1{r^2}\left(\frac1{2e^2r^2}+\eta^2\right), \ T_\theta^\theta
=T^\phi_\phi=\frac1{2e^2r^4} \ .
\end{equation}
The gravitational field associated with the above tensor reads:
\begin{equation}
B(r)=\frac1{A(r)}=\alpha^2-\frac{2GM}r+\frac{4\pi G}{e^2r^2} \ ,
\end{equation} 
which corresponds to the Reissner-Nordstr\"{o}m spacetime with an additional
solid angle deficit factor. This metric tensor, as mentioned above, describes 
the effect produced in the geometry by two distinct objects: the 
global monopole, responsible for a solid angle deficit, and the magnetic 
monopoles, responsible for a non-vanishing radial magnetic field
\begin{equation}
B_i=B_i^a\hat{\phi}^a=\hat{x}_i/er^2 \ .
\end{equation}

Although the above expressions represent exact solutions for this system, 
unfortunately they cannot be accepted as physical solution. The non-integrable
factor $1/r^4$ of $T^t_t$ provides an infinity energy inside
a finite space region around the defect. Finally we want to say that 
the scalar curvature associated with the above spacetime is 
$R=\frac{2(1-\alpha^2)}{r^2}$.

\section{Numerical Analysis}

In this section we shall exhibit the most relevant aspects about this
compost defect under a numerical analysis. Our strategy is to present
numerical solutions for the matter and gravitational fields which obey 
regularity conditions at origin. See Eq. (\ref{O}). Mainly we are interested to
analyse their behaviors as the parameters $\eta$, associated with the energy
scale where the symmetry is spontaneously broken and $\lambda$, the 
self-coupling constant, both vary. In order to start the numerical analysis
we shall express the set of differential equations, (\ref{AB}) - (\ref{F}), 
in terms of two dimensionless parameters $\Delta=8\pi G\eta^2$ and $\beta=
\lambda/e^2$, rescaling the radial coordinate $r$ as $x=re\eta$. 

The case $\Delta=0$, i.e., $G=0$ corresponds to the flat-space one. The 
solutions for the matter fields is the 't Hooft-Polyakov magnetic monopole
for the local sector with the global sector independent. Choosing $\beta=0$ 
the solution for the $u$ and $h$ can be given in a closed form \cite{PS}.
As to the global sector, vanishing $\beta$ the system does not
provide localized solution: the differential equation for $f$ becomes
linear and a regular solution at origin diverges as $r \to \infty$. An
intermediate situation happens when we assume $\Delta \neq 0$ and the
self-coupling constant for the local Higgs sector only vanishes. In
this case the matter field equation becomes
\begin{equation}
\label{AB1}
\frac1{\sqrt{AB}}\left(\sqrt{\frac BA}\ u'\right)'=\frac{u(u^2-1)}{x^2}+
h^2u \ ,
\end{equation}
\begin{equation}
\frac1{x^2 \sqrt{AB}}\left(x^2 \sqrt{\frac BA}\ h'\right)'=\frac{2hu^2}{x^2}
\end{equation}
and
\begin{equation}
\label{F1}
\frac1{x^2\sqrt{AB}}\left(x^2\sqrt{\frac BA} \ f'\right)'=\frac{2f}{x^2}+
\beta f(f^2-1) \ ,
\end{equation}
where the primes in the above equations denote differentiation with
respect to $x$.

As we have said before, in this section we shall analyse, numerically,
both cases: The first one described by Eqs. (\ref{AB}) - (\ref{F}), and 
the second one by Eqs. (\ref{AB1}) - (\ref{F1}).

Now let us start first with the complete model. Casting the differential 
equations in first-order form by auxiliaries fields $P=u'$, $Q=h'$, $D=f'$ and 
defining a new other variable $g=1/A$, the set of differential equations 
becomes
\begin{equation}
P=u', \ Q=h', \ D=f' ,  \ g=1/A \ ,
\end{equation}
\begin{equation}
g'=\frac1x-\frac gx-\Delta x\left[U(f,h,u)+g\left(\frac{P^2}
{x^2}+\frac{Q^2}2+\frac{D^2}2\right)\right]\ ,
\end{equation}
\begin{equation}
P'=\frac1g\left[\frac{gP}x-\frac Px+\frac{u(u^2-1)}{x^2}+h^2u+\Delta xP\ 
U(f,h,u)
\right] \ ,
\end{equation}
\begin{equation}
\label{Q}
Q'=\frac1g\left[\frac{2hu^2}{x^2}+\beta h(h^2-1)-\frac{gQ}x-\frac Qx+
\Delta xQ U(f,h,u)\right] \ 
\end{equation}
and
\begin{equation}
D'=\frac1g\left[\frac{2f}{x^2}+\beta f(f^2-1)-\frac{gD}x-\frac Dx+
\Delta xD U(f,h,u)\right] \ ,
\end{equation}
being
\begin{equation}
\label{U1}
U(f,h,u)=\frac{(u^2-1)^2}{2x^4}+\frac{u^2f^2}{x^2}+\frac{f^2}{x^2}+
\frac\beta 4(h^2-1)^2+\frac\beta 4(f^2-1)^2 \ .
\end{equation}

Near the origin, regular solutions must behave as
\begin{equation}
\label{f}
f=c_fx+O(x^3), \ h=c_hx+O(x^3), \ u=1-c_ux^2+O(x^4), \ 
\end{equation}
and
\begin{equation}
g=1-\Delta\left[2c_u+\frac12\left(c_f^2+c_h^2\right)+\frac\beta6\right]x^2+
O(x^4) \ ,
\end{equation}
where the three constants $c_f$, $c_h$ and $c_u$ must be chosen in order 
to have $f$, $h$ and $u$ approaching to the correct values as 
$x\rightarrow\infty$.

The case $c_h=c_u=0$ and $c_f\neq 0$ corresponds to the global monopole
spacetime. In this case there is only one constant to be adjusted. The set
of differential equations presents only one parameter $\Delta$. This model
has been first numerically analyzed by Harari and Loust\'{o} \cite{Lousto}. 
There, they show that the behavior of the Higgs field is quite insensitive
to the values of $\Delta$ in the interval $0\leq \Delta \leq 1$. More recently,
Maison and Liebling \cite{Maison} and Liebling \cite{Liebling} returned to 
the numerical analysis of this model and found that for $\Delta \geq 1$, 
$1/A$ decreases toward zero indicating the presence of a horizon.

A more complicated case is when $c_f=0$ with $c_h$ and $c_u$ different from 
zero. This case corresponds to a gravitating magnetic monopole. There are 
two constants to be adjusted numerically in order the system to present
localized topological defect. This model has been analyzed by Lee {\it et
al} \cite{Lee}, Ortiz \cite{Ortiz} and  Breitenlohner {\it et al} 
\cite{Peter}. In these papers the authors observed that the system presents 
singular solutions when $\Delta$ is greater than some critical value, 
$\Delta_{cr}$. For these situations $1/A$ has zeros, the Schwarzchild 
radius becomes greater than the monopole's size, so the monopole must be 
a black-hole.

Now returning to our system, we present in what follows our numerical results.
Defining by the radius of the global and magnetic monopoles' core the value 
of the dimensionless variable $x$ corresponding to $f(x_L)=0.9$ and 
$h(x_G)=0.9$, respectively, we can observe by Figs. $1(a)$ and $1(b)$, 
that $r_L<r_G$. Also we can notice that both radiuses decrease as $\beta$ 
becomes larger. So, these results confirm that, for this model, the magnetic 
monopole configuration 
approaches to its vacuum value faster than the global monopole. In this sense 
the magnetic monopole's core is firstly formed. Moreover, other graphs
not included in this paper indicate that the shapes of $f(x)$ and $h(x)$ are 
almost insensitive to the values of the parameter $\Delta$.\footnote{A similar 
conclusion has been reached by Harari and Loust\'{o} \cite{Lousto} in their 
analysis of a pure global monopole system}

The Figs. $2(a)$ and $2(b)$ display the behavior of the function $u(x)$ 
with $x$ for different values of $\beta$ and $\Delta$. From them it is 
possible to observe that its behavior is sensitive to the values of the
parameters $\Delta$ and $\beta$, in such way that $u$ reaches its asymptotic 
value faster for greater values of these two parameters.

The Figs. $3(a)$ and $3(b)$ exhibit the behavior of the fields $f$ and $h$
for fixed value of $\Delta$ and different values of $\beta$. We can see that
their behaviors are very sensitive to this parameter, and that their
radius decrease when $\beta$ increases. From numerical point of view, 
solutions with large $\beta$ become more difficult to be analysed, this is
the reason why they are presented in different intervals of
the variable $x$. \footnote{The stability problem related with numerical
solutions for large $\beta$ has been pointed by Breitenlohner {\it at all} 
in Ref. \cite{Peter} for $\beta>5$ in the gravitating magnetic monopole
system.} 

As to $g(x)=\frac1{A(x)}$, which asymptotes to non-unity values $\alpha^2=
1-\Delta$, it develops a local minimum for large values of the parameters
$\Delta$ and $\beta$, independently. Moreover, as $\Delta$ increases
the asymptotic value of $g$ decreases toward zero, and becomes negative
for $\Delta > 1$ indicating the presence of a horizon. So, for $\Delta \geq 1$
this system presents a horizon for any nonzero value of $\beta$. However, 
for $\Delta <1$ there exist a critical value for $\beta$ above which this
compost defect becomes a black hole. To find a domain of existence of
regular solution is possible only formally, analysing the set of
differential equations at horizon, i. e., substituting $g=0$ at the
point $x=x_H$ in the set. Only numerical calculations allows the obtainement 
of related parameters $\beta$ and $\Delta$ associated with a specific
singular solution. 

The figures $4(a)$ and $4(b)$ exhibit the behavior of $M(x)$ with $x$, 
being $M(x)$ a dimensionless function obtained by $M(r)$ given in (\ref{M}). 
In fact this dimensionless mass function which depends only the two
parameter $\beta$ and $\Delta$ is defined by $M(r)=4\pi\eta/e M(x)$.
The asymptotic behavior of $M(x)$ provides information about the effective 
monopole mass. Fig $4(a)$ shows that for $\beta=1$ this function 
asymptotes a negative value. This very peculiar feature has been detected 
for the global monopole defect by Harari and Loust\'{o} in \cite{Lousto}. 
\footnote{In \cite{Lousto} it was observed that the shape of the curves are 
very insensitive to $\Delta$ in the interval $0\leq\Delta\leq 1$.} However for 
$\beta=10$ the Fig. $4(b)$ shows that the effective mass of this topological 
defect becomes positive. (The same behavior is observed for $\beta=80$.) So 
this compost defect presents repulsive or attractive gravitational 
interactions which depends on the magnitude of the self-coupling constant 
$\lambda$.

The second case can be numerically analyzed in similar way as the
previous one; however some changes must be done in order to take
into account the vanishing of the self-coupling constant in the local
sector of the system. The first-order differential equation set for 
this case can be written discarding the terms $h(h^2-1)$ in (\ref{Q})
and (\ref{U1}).  The behavior for the fields $f$, $h$ and $u$ at the origin
are similar to (\ref{f}); however for $g$ it is
\begin{equation}
g=1-\Delta\left[2c_u+\frac12\left(c_f^2+c_h^2\right)+\frac\beta{12}\right]x^2+
O(x^4) \ .
\end{equation}
Once more three new constants must be chosen in order to have solutions with 
appropriated behavior at infinity.

The most important characteristic observed by us about this model are 
summarized below:\\
$i)$ The Figs. $5(a)$ and $5(b)$ show the behavior of the fields $f$
and $h$. Considering again the same definition to the radius of the 
defects as given before, we can see that for this case $r_G<r_L$. This is 
in contrast with the result found in the previous model. So comparing 
the results found in these two models it is possible to conclude that 
the sizes of the global and magnetic monopole's core depends on the 
intensity of their respective self-interactions. Moreover, we can
infer that for specific values of these constants, both topological defects 
present equal radius, though we cannot ensure that both fields $f$ and $h$
have the same behavior.\\
$ii)$ The Figs. $6(a)$ and $6(b)$ exhibit, respectively, an explicit 
dependence of $u$ and $h$ with $\beta$. Although there is no direct 
interaction between the local sector, represented by these fields, with
the self-coupling constant associated with the global sector, our
numerical analysis indicate a sensitive dependence of both fields
with $\lambda$.\\
$iii)$ The Fig. $6(c)$ exhibits the behavior of $f$ with $\beta$. In 
this case its dependence is more prominent than for $h$, i.e., the radius
of the global monopole decreases more rapidly with the increasing of
$\beta$ than the magnetic's one.

As to the effective mass associated with this case, $M(x)$, it is observed 
the same behavior exhibited in the previous case. So we decided do not included
extra figures in this part.

\section{Concluding Remarks}

In this paper we have presented a model which describes two topological
defects at the same time: The global and magnetic monopole in a curved
spacetime. The Lagrangian density which governs this system contains
two distinct bosonic sectors. In order to make our analysis easier we
decided do not include a direct interaction between them. Two different
situations were analyzed: the first one considering the Higgs 
self-interactions and vacuum expectation values equal for both sectors. 
The second situation is a particular case when we switch off the 
self-coupling constant associated with the local sector only. In both
cases the set of coupled differential equations does not allow to
obtain a closed solutions, even singular. Only asymptotic behavior
for matter and gravitational fields can be provided analytically.
Specifically, for regions very far from the topological defect's
core, the spacetime corresponds to a  Reissner-Nordstr\"{o}m spacetime
with a solid angle deficit factor
\begin{eqnarray}
B(r)=\frac1{A(r)}&=&\alpha^2-\frac{2GM}r+\frac{4\pi G}{e^2r^2} \ . \nonumber
\end{eqnarray}

Here we have provided numerical informations about the behavior 
of these fields in a non-asymptotic region. These informations concern to
the relative sizes of both defects, their dependence on the two parameters 
presented in this model, the self-coupling constant, through $\beta$, and the 
gravitational constant, through $\Delta$, etc. 

The numerical method applied by us in this paper was double-precision
fourth-order Runge-Kutta routine. For all calculations the errors found 
were of order $10^{-3}$ or less.

It is our intention to continue investigating the behavior of the fields 
for larger value of the parameter $\Delta$. As shown in previous 
papers analysing global \cite{Maison, Liebling} and gravitational 
\cite{Lee, Nieuw, Peter} monopoles, for $\Delta$ bigger than some critical 
value, the system presents horizons. For both distinct cases, the horizons
appears when $\Delta$ is of order unity.

\section{Figure Captions}

{\bf Figure 1}: These graphs show simultaneously the behavior of $f$ and $h$ 
for: $(a)$ $\beta=1$ and $\Delta=0.1$ and $(b)$ for $\beta=80$ and 
$\Delta=10^{-6}$.\hfill\\
[6mm]
{\bf Figure 2}: These graphs show the behavior of $u$ for three different
values of $\Delta$ for $(a)$ $\beta=10$ and $(b)$ $\beta=80$.\hfill\\[6mm]
{\bf Figure 3}: These graphs show the behavior of $(a)$ $f$ and $(b)$ $h$ 
for $\Delta=0.1$ and three different values of $\beta$.\hfill\\
[6mm]
{\bf Figure 4}: These graphs show the behavior of the effective mass, $M(x)$
for $(a)$ $\beta=1$ and $(b)$ $\beta=10$ for three different values of 
$\Delta$.\hfill\\
[6mm]
{\bf Figure 5}: These graphs show simultaneously the behavior of $f$ and $h$ 
for: $(a)$ $\beta=10$ and $\Delta=10^{-6}$ and $(b)$ for $\beta=10$ and 
$\Delta=0.1$.\hfill\\
[6mm]
{\bf Figure 6}: These graphs show the behavior of $(a)$ $u$, $(b)$ $h$ and
$(c)$ $f$ for three different values of $\beta$ for $\Delta=0.1$.\hfill\\
[6mm]

\begin{figure}[t]
\begin{center}
\includegraphics[width=6cm,angle=-90]{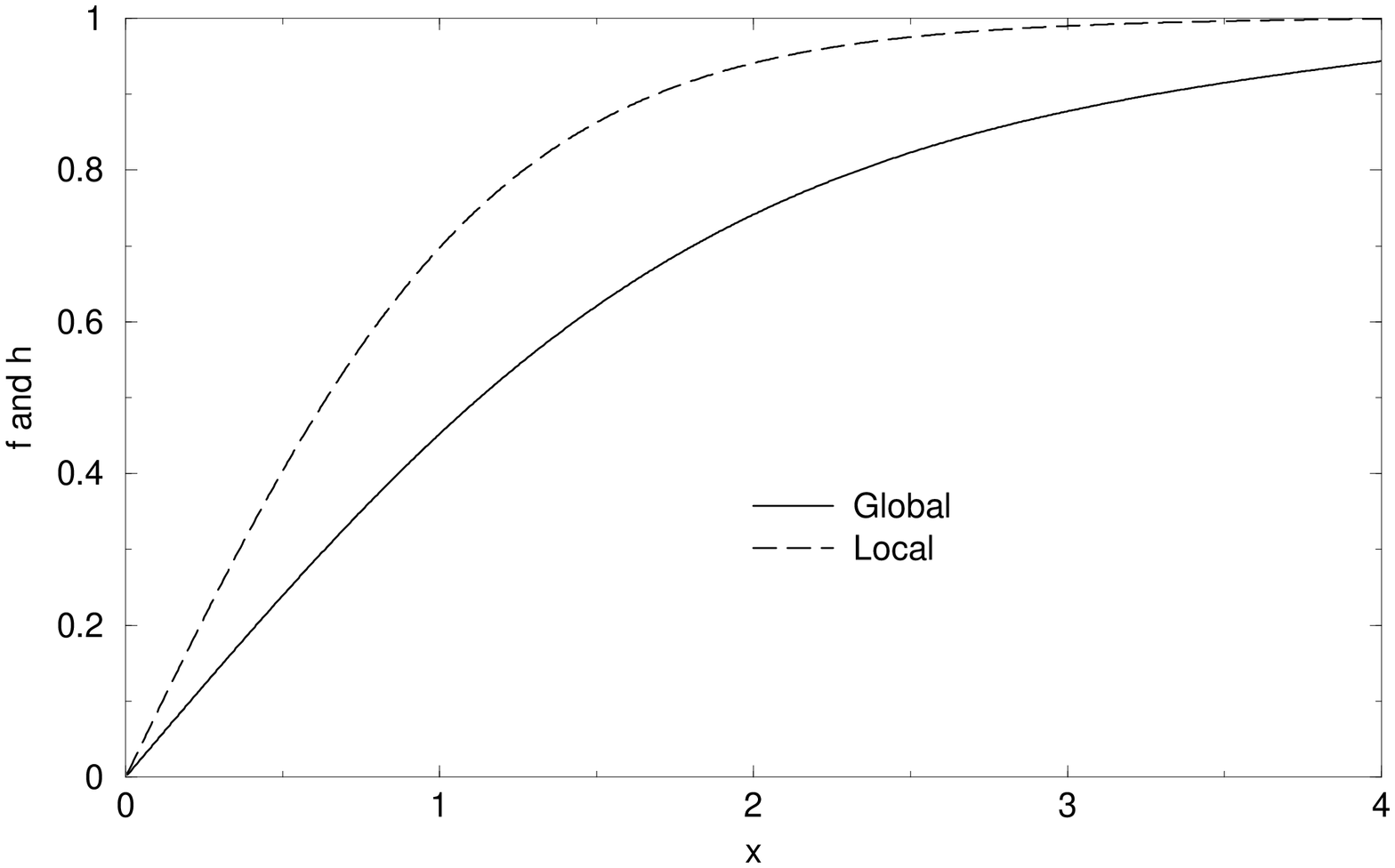}

(a)

\includegraphics[width=6cm,angle=-90]{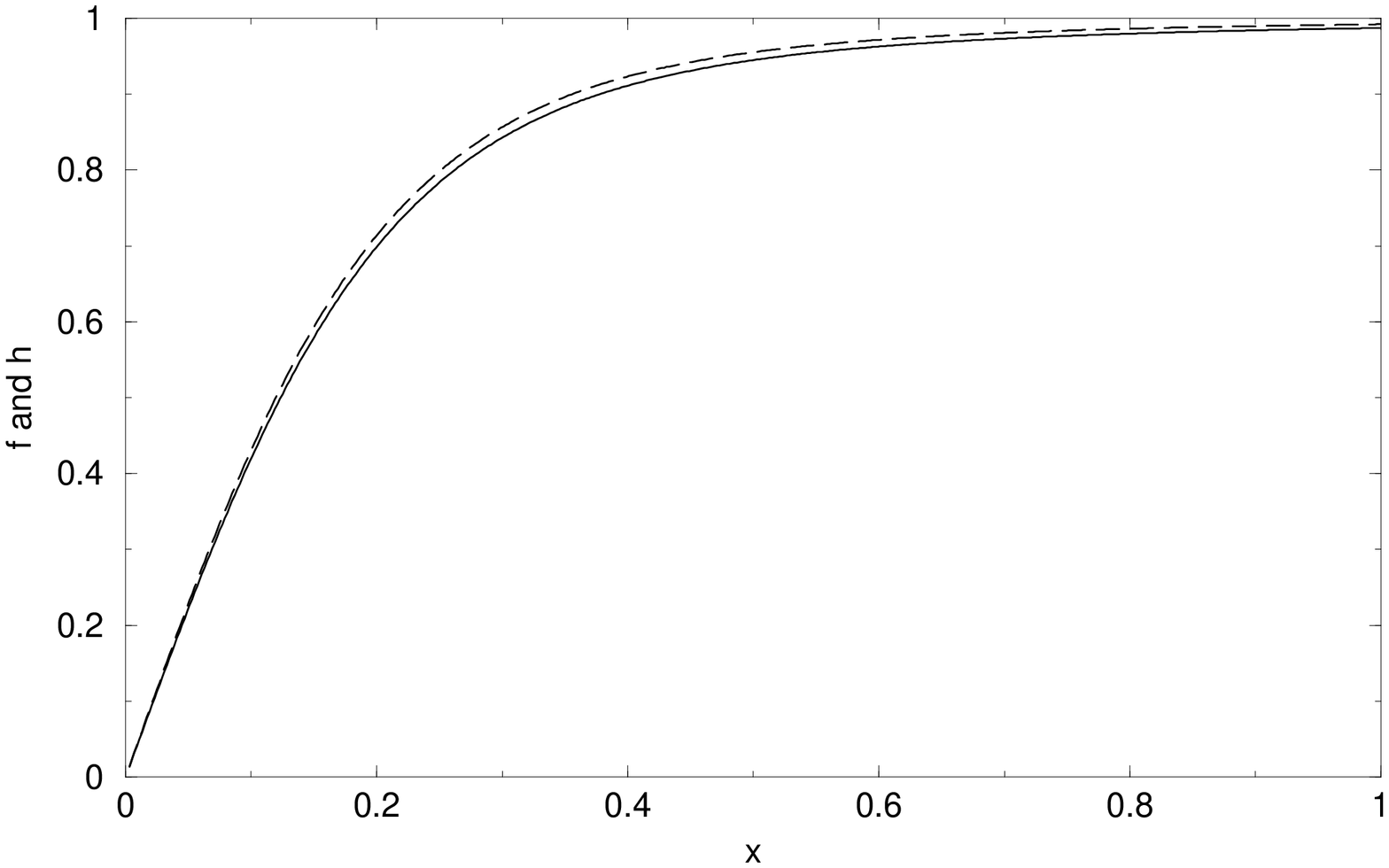}

(b)

\label{fig1}
\caption{}
\end{center}
\end{figure}

\newpage

\begin{figure}[t]
\begin{center}
\includegraphics[width=6cm,angle=-90]{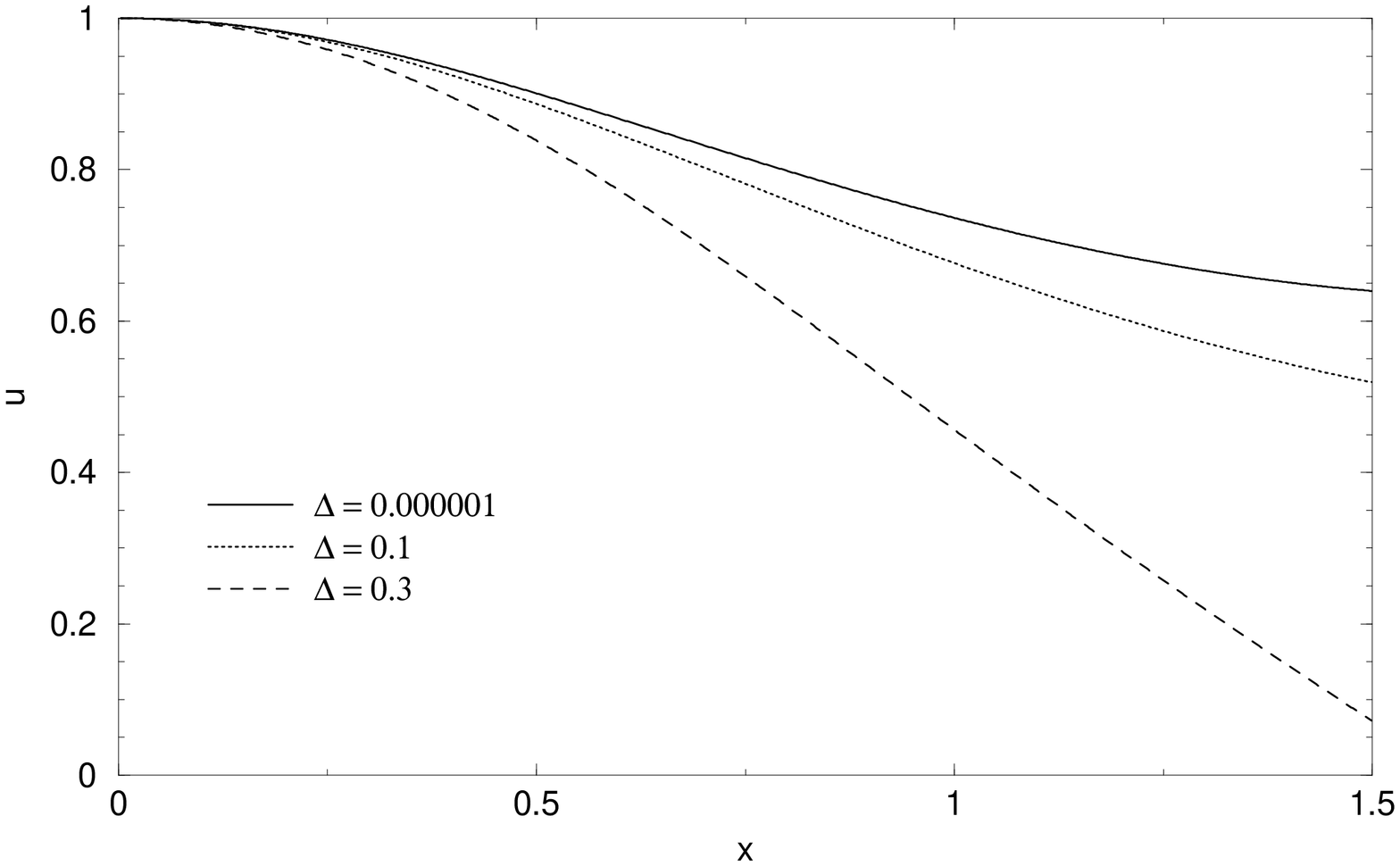}

(a)

\includegraphics[width=6cm,angle=-90]{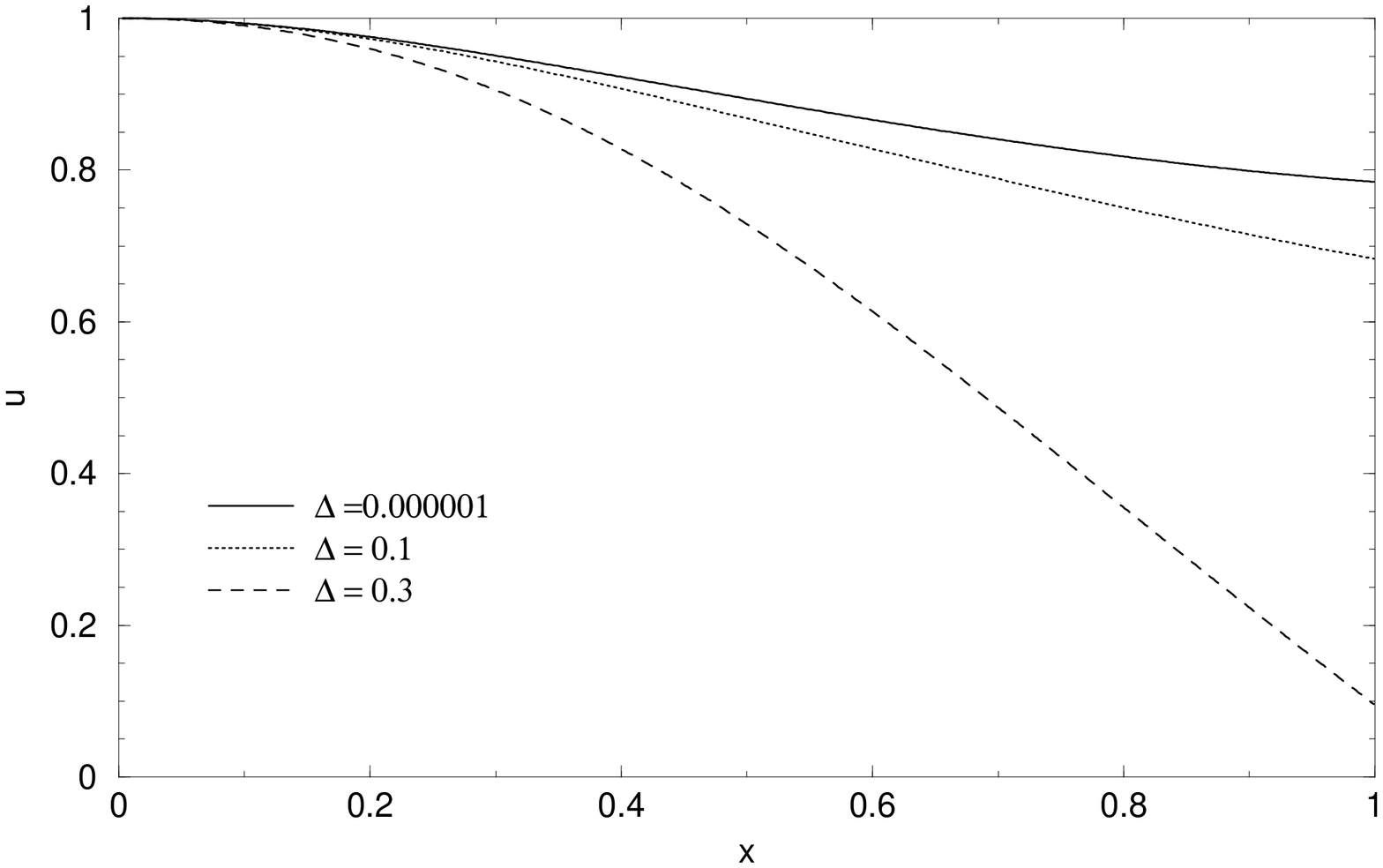}

(b)

\label{fig2}
\caption{}
\end{center}
\end{figure}

\newpage

\begin{figure}[t]
\begin{center}
\includegraphics[width=6cm,angle=-90]{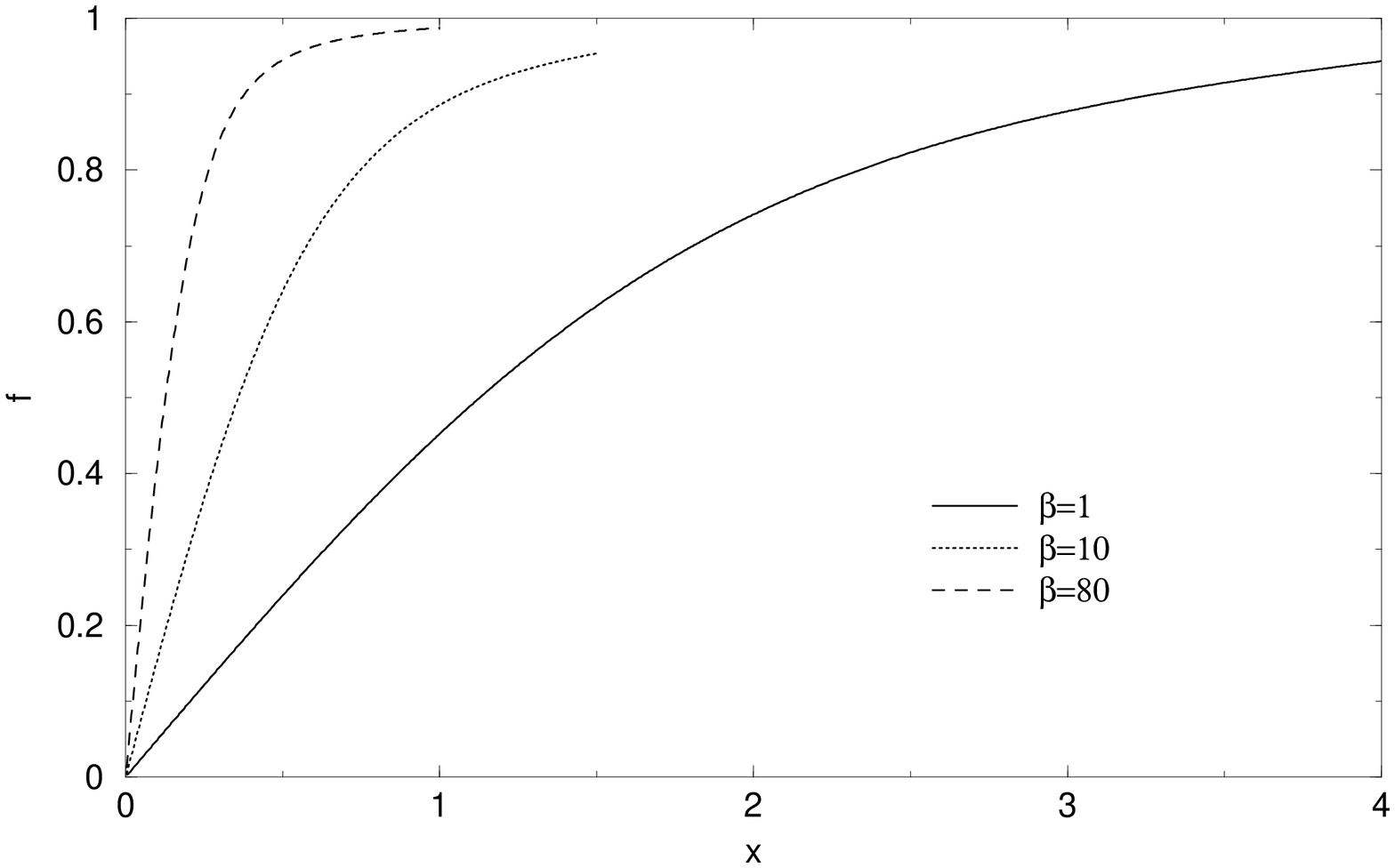}

(a)

\includegraphics[width=6cm,angle=-90]{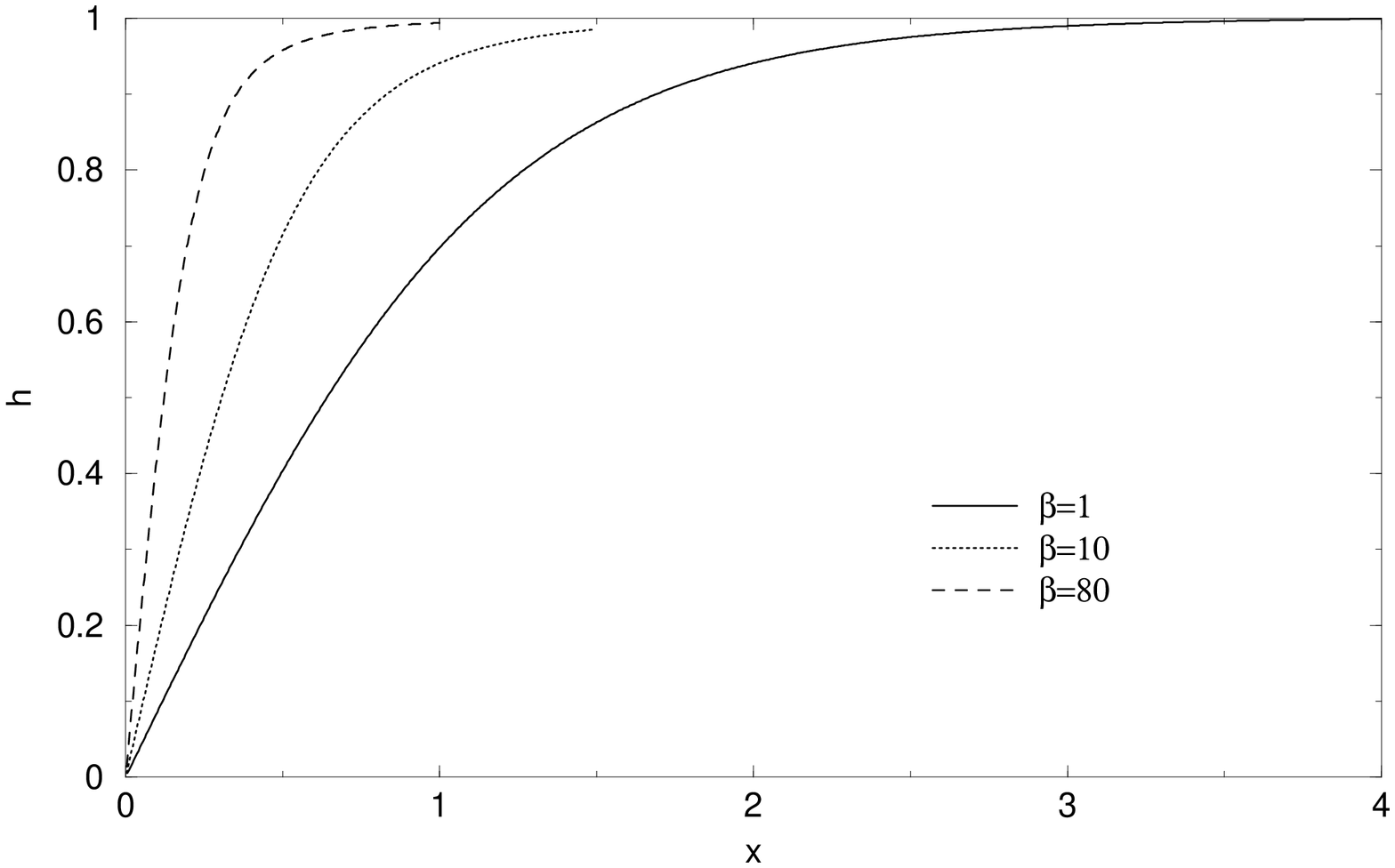}

(b)

\label{fig3}
\caption{}
\end{center}
\end{figure}

\newpage

\begin{figure}[t]
\begin{center}
\includegraphics[width=6cm,angle=-90]{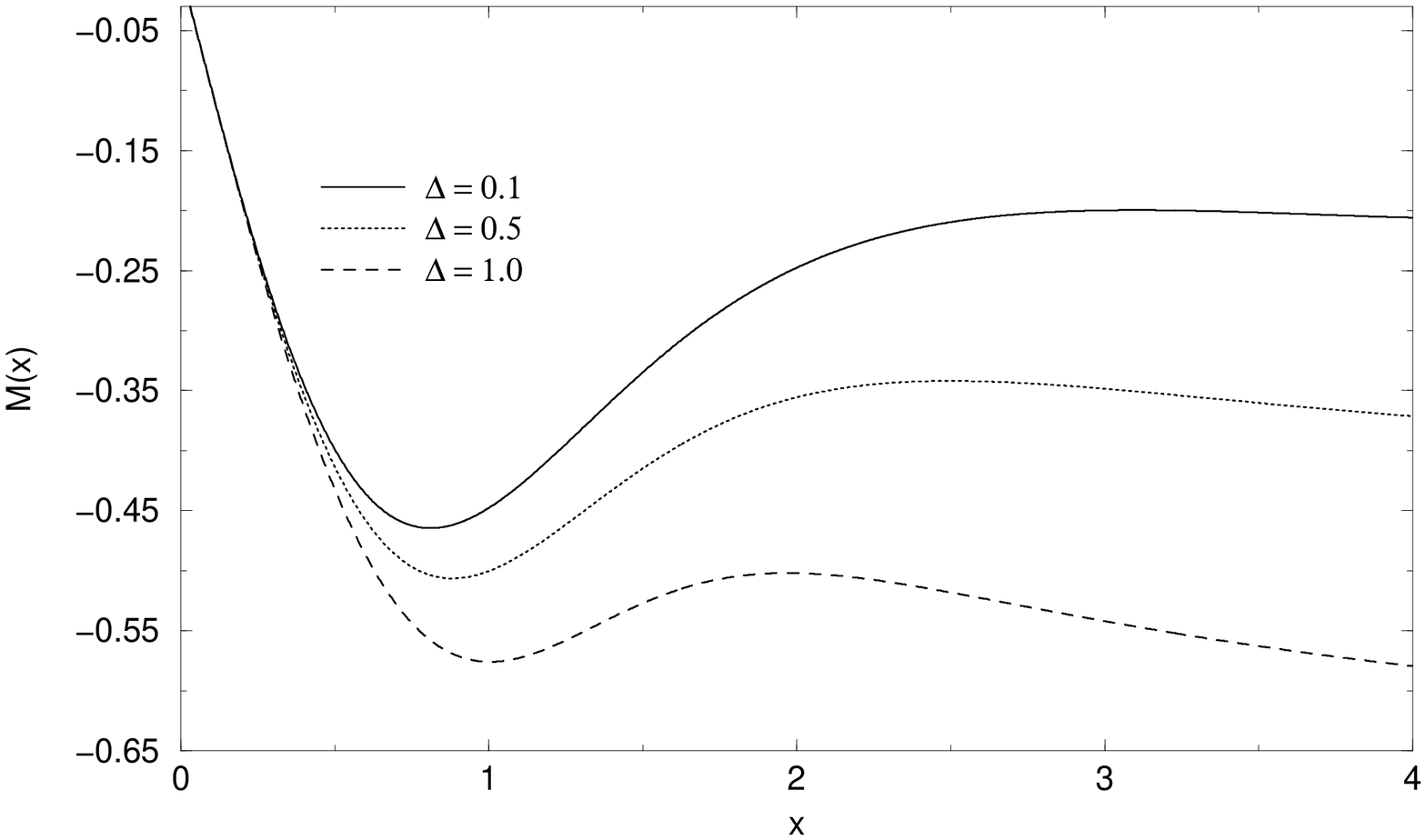}

(a)

\includegraphics[width=6cm,angle=-90]{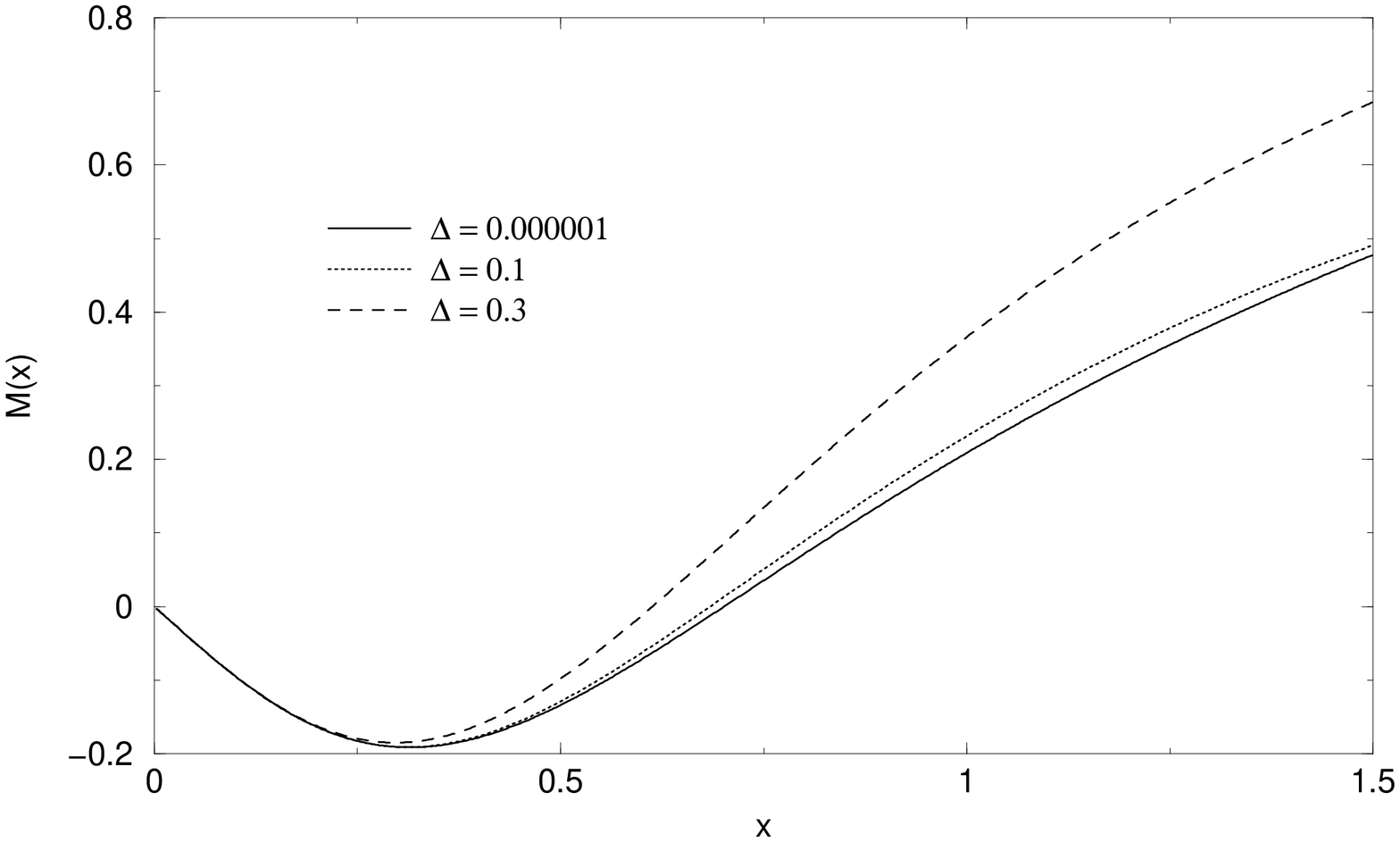}

(b)

\label{fig4}
\caption{}
\end{center}
\end{figure}

\newpage

\begin{figure}[t]
\begin{center}
\includegraphics[width=6cm,angle=-90]{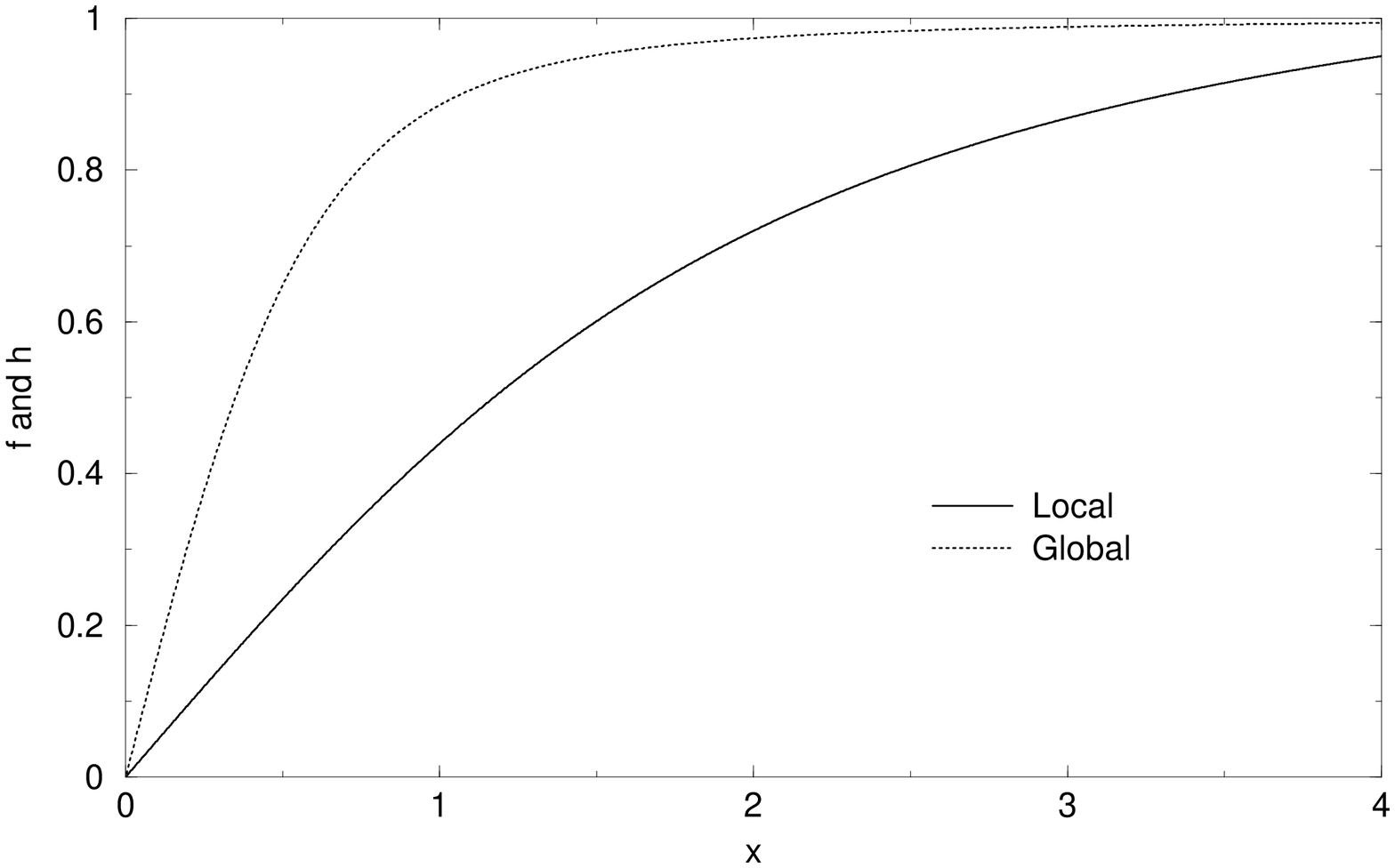}

(a)

\includegraphics[width=6cm,angle=-90]{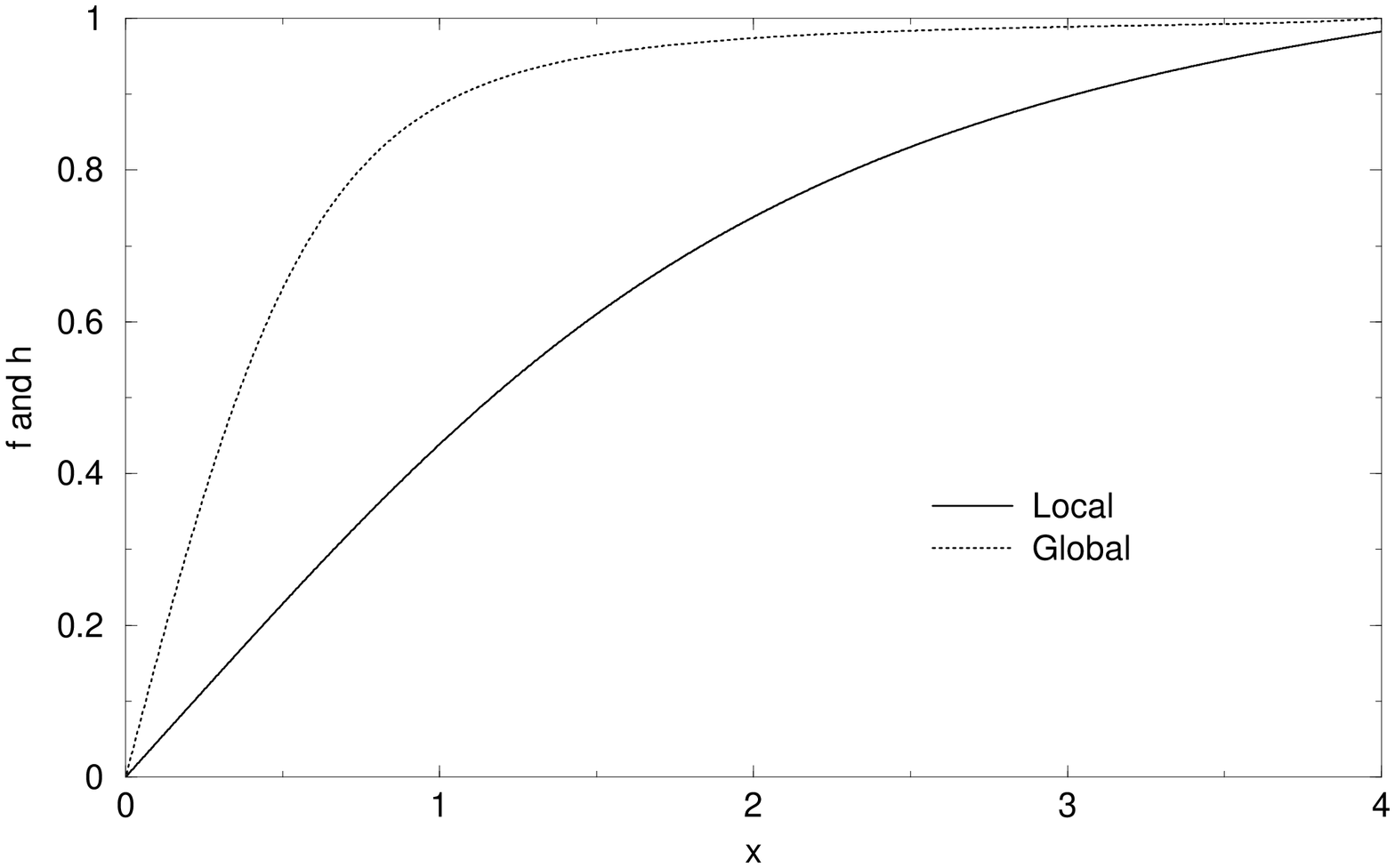}

(b)

\label{fig5}
\caption{}
\end{center}
\end{figure}

\newpage

\begin{figure}[t]
\begin{center}
\includegraphics[width=5.1cm,angle=-90]{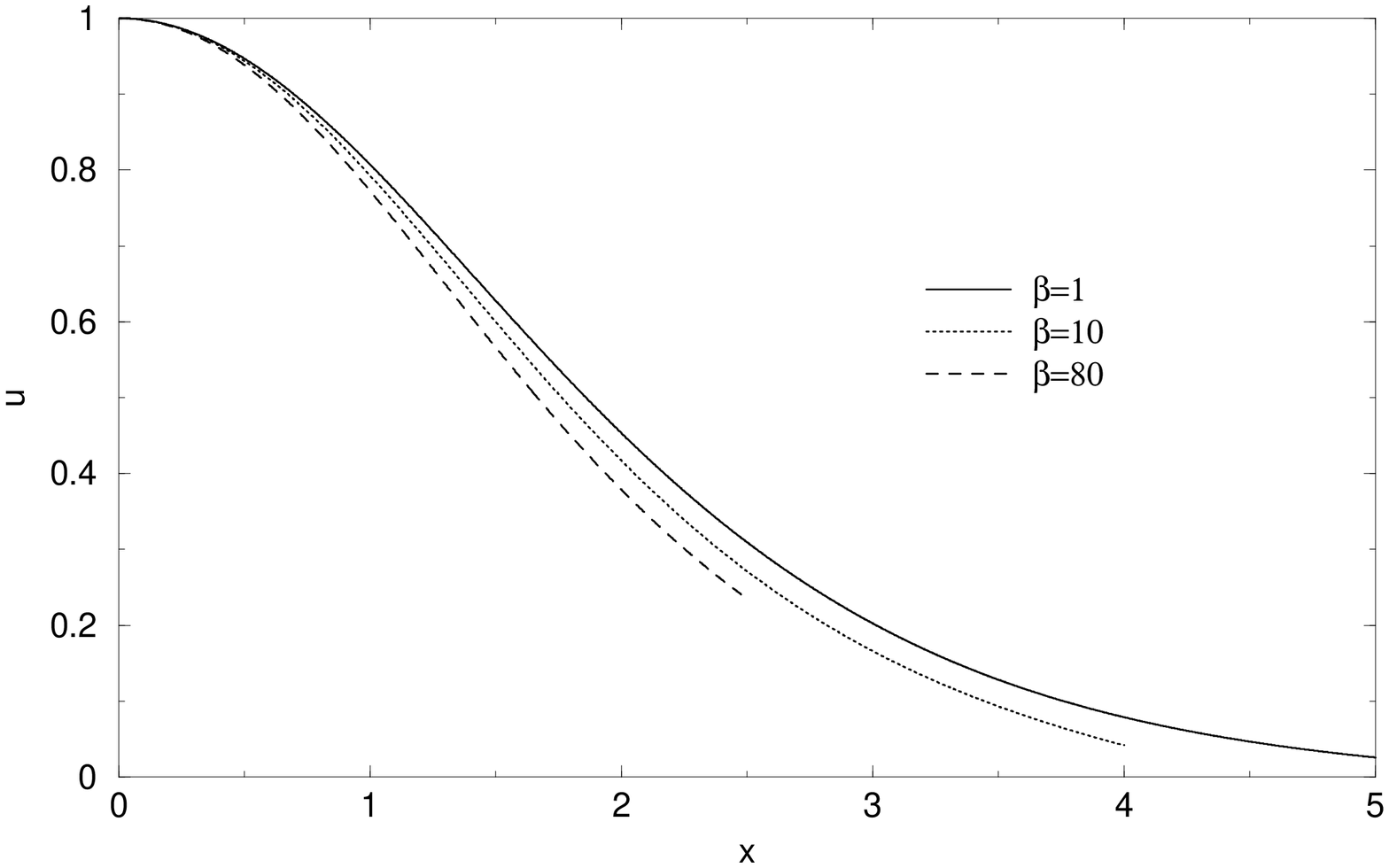}

(a)

\includegraphics[width=5.1cm,angle=-90]{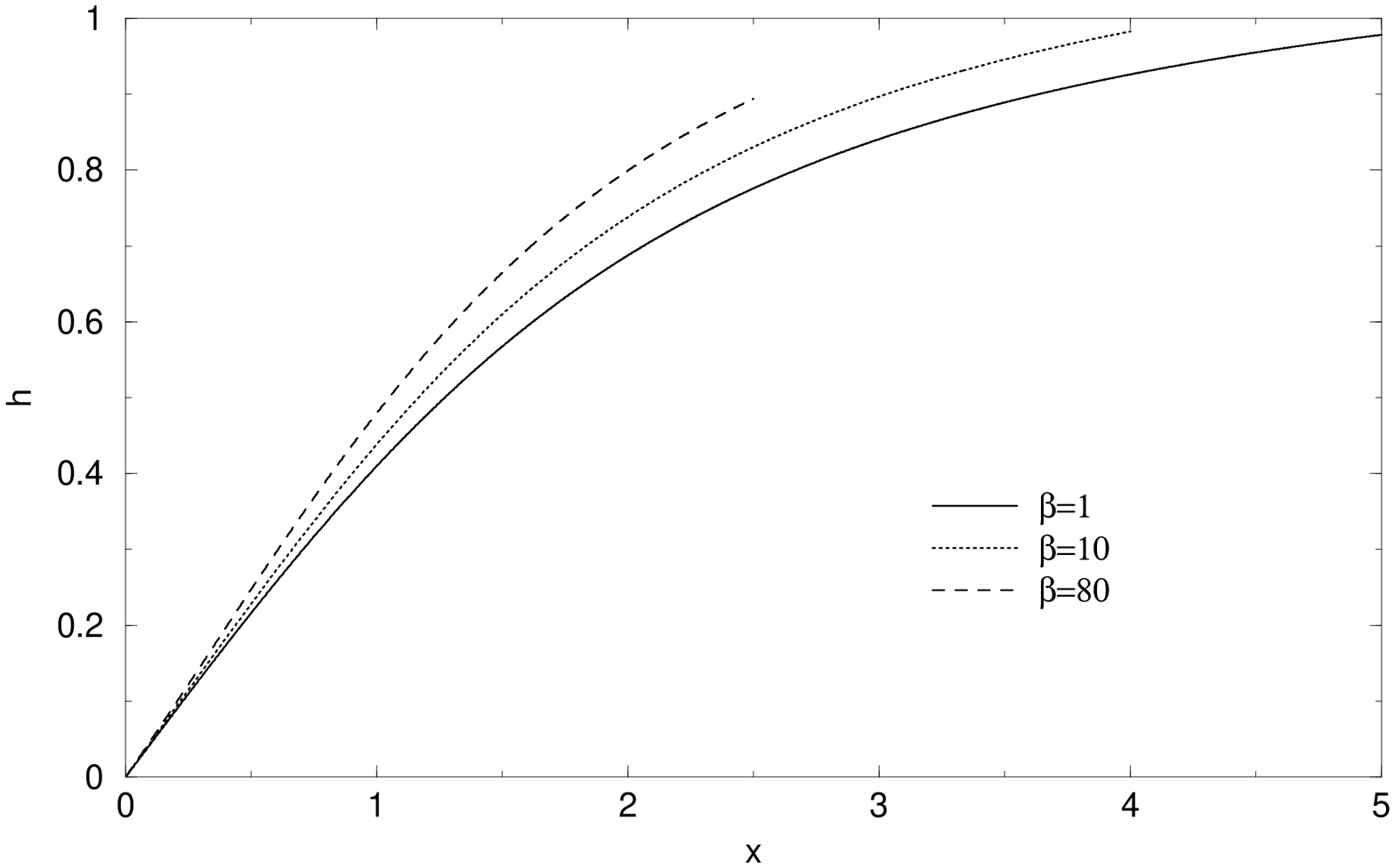}

(b)

\includegraphics[width=5.1cm,angle=-90]{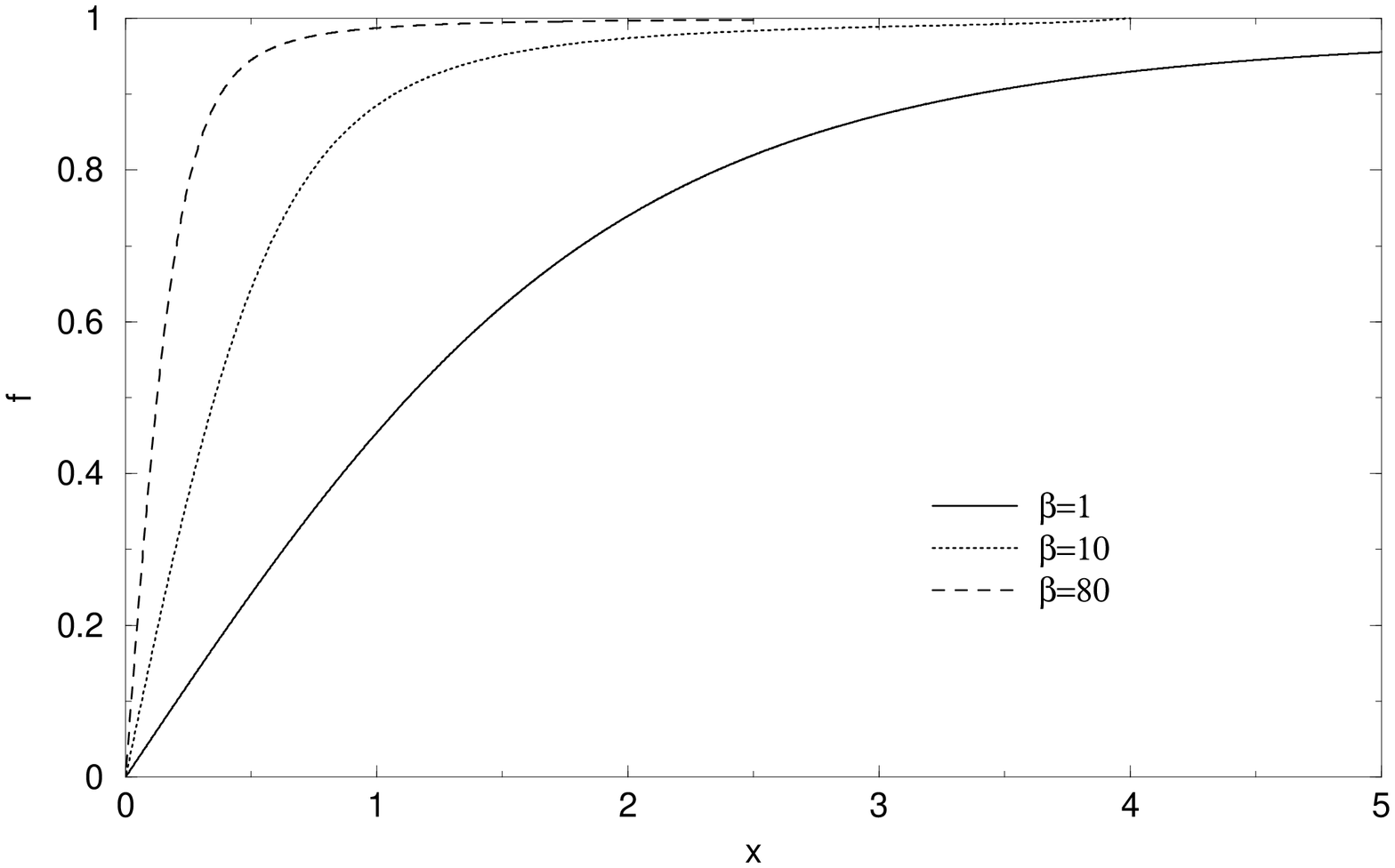}

(c)
\label{fig6}
\caption{}
\end{center}
\end{figure}


\newpage
\begin{thebibliography}{100}
\bibitem{HP} G. 't Hooft, Nucl. Phys. {\bf B79}, 276 (1974) and A. M. Polyakov,
Pisma v. Zh. E.T.F. {\bf 20}, 430 (1974), (JETP Lett. {\bf 20}, 194 (1974)).
\bibitem{Bais} F. A. Bais and R. J. Russel, Phys. Rev. D {\bf 11}, 2692 (1975).
\bibitem{Cho}Y. M. Cho and P. G. O. Freund, Phys. Rev. D {\bf 12}, 1588 (1975).
\bibitem{Nieuw} P. van Nieuwenhuizen, D. Wilkinson and M. J. Perry, Phys.
Rev. D {\bf 13}, 778 (1976).
\bibitem{Lee} K. Lee, V. P. Nair and E. Weinberg, Phys. Rev. D {\bf 45}, 2751
(1992).
\bibitem{Ortiz} M. E. Ortiz, Phys. Rev. D {\bf 45} R2586 (1992).
\bibitem{BV} M. Barriola and A. Vilenkin, Phys. Rev. Lett. {\bf 63}, 341 
(1989).
\bibitem{Salgado} U. Nucamendi, M Salgado and D. Sudarsky, Phys. Rev. Lett.
{\bf 84}, 3037 (2000).
\bibitem{Parker} M. S. Turner, E. N. Parker and T. J. Bogdan, Phys. Rev. D
{\bf 26}, 1296 (1982).
\bibitem{Hiscock} W. Hiscock, Phys. Rev. Lett. {\bf 64}, 344 (1990).
\bibitem{BE} D. P. Bennett and S. H. Rhie, Phys. Rev. Lett. {\bf 65}, 1709
(1990).
\bibitem{Peter} P. Breitenlohner, P. Forg\'acs and D. Maison, Nucl. Phys.
{\bf B383}, 357 (1992) and  P. Breitenlohner, P. Forg\'acs and D. Maison,
Nucl. Phys. {\bf B442}, 126 (1995).
\bibitem{Lousto} D. Harari and C. Loust\'{o}, Phys. Rev. D {\bf 42}, 2626 
(1990).
\bibitem{Maison} D. Maison and S. Liebling, Phys. Rev. Lett. {\bf 83}, 5218 
(1999).
\bibitem{Liebling} S. L. Liebling, Phys. Rev. D {\bf 61}, 024030 (2000).
\bibitem{PS} M. K. Prasad and C. M. Sommerfeld, Phys. Rev. Lett. {\bf 35}, 760
(1975).

\end{thebibliography}
\end{document}